\documentclass[10pt,conference]{IEEEtran}

\usepackage{cite}
\usepackage{amsmath,amssymb,amsfonts}
\usepackage{algorithmic}
\usepackage{graphicx}
\usepackage{textcomp}
\usepackage{xcolor}
\usepackage{enumitem}
\usepackage{amsmath}
\usepackage{bbold}
\usepackage{comment}
\usepackage[utf8]{inputenc}
\usepackage{graphicx}
\usepackage{hyperref}
\usepackage{verbatim}
\usepackage{cleveref}
\usepackage{listings}
\usepackage{booktabs}
\usepackage{cite}
\usepackage{subfigure}
\usepackage{framed}

\def\BibTeX{{\rm B\kern-.05em{\sc i\kern-.025em b}\kern-.08em
    T\kern-.1667em\lower.7ex\hbox{E}\kern-.125emX}}

\graphicspath{{./figs/}}

\title{Actionable Mutation Testing at Scale}
\title{Combining Mutation Testing and Reverse Test Coverage to Make Mutants Actionable}
\title{Highly Covered, yet Poorly Tested: Making Mutation Testing Actionable at Scale with Mutation Monkey}
\title{Toward More Actionable and Scalable Mutation Testing at Facebook}
\title{What It Would Take to Use Mutation Testing in Industry---A Study at Facebook}

\author{\IEEEauthorblockN{Moritz Beller,\IEEEauthorrefmark{1} Chu-Pan Wong,\IEEEauthorrefmark{3} Johannes Bader,\IEEEauthorrefmark{4} Andrew Scott,\IEEEauthorrefmark{1} Mateusz Machalica,\IEEEauthorrefmark{1} Satish Chandra,\IEEEauthorrefmark{1} Erik Meijer\IEEEauthorrefmark{1}}\vspace{0.25cm}
\IEEEauthorblockA{\IEEEauthorrefmark{1}\textit{Probability},
\textit{Facebook, Inc.},
Menlo Park, USA\\
\{mmb, andrewscott, msm, satch, erikm\}@fb.com} \vspace{0.15cm}
\IEEEauthorblockA{\IEEEauthorrefmark{3}\textit{Carnegie Mellon University (work done while interning at Facebook)},
Pittsburgh, USA \\
chupanw@cs.cmu.edu}\vspace{0.15cm}
\IEEEauthorblockA{\IEEEauthorrefmark{4}\textit{Jane Street Capital (ex-employee; work done while at Facebook)}\\
johannes-bader@hotmail.de}
}
\vspace{-0.8cm}
\begin{document}
\maketitle

\begin{abstract}
Traditionally, mutation
testing generates an abundance of small deviations of a program, called
mutants. At industrial systems the scale and size of Facebook's, doing this is infeasible. We should not create mutants that the test suite would likely fail on or that give no actionable signal to
developers. To tackle this problem, in this paper, we semi-automatically learn
error-inducing patterns from a corpus of common Java coding errors and from changes that caused operational anomalies at Facebook
specifically. We combine the mutations with instrumentation that measures which tests exactly visited the mutated piece of code. Results on more than 15,000 generated mutants show that  more than half of the generated mutants survive Facebook's rigorous test suite of unit, integration, and system tests. Moreover, in a case study with 26 developers, all but two expressed that the mutation exposed a lack of testing in principle. As such, almost half of the 26 would actually act on the mutant presented to them by adapting an existing or creating a new test. The others did not for a variety of reasons often outside the scope of mutation testing. It remains a practical challenge how we can include such external information to increase the actionability rate on mutants.
\end{abstract}

\begin{IEEEkeywords}
Mutation Testing, Machine Learning, Getafix, Mutation Monkey
\end{IEEEkeywords}

\maketitle

\section{Introduction}

Mutation testing is a way to assess the quality of the test suite of a
software program~\cite{demillo1978hints}. Traditionally, mutation
testing generates an abundance of small deviations of a program, called
mutants. Mutants are generated by applying a mutation operator on the original
program---e.g., deleting a method call, disabling an {\tt if}
condition, or replacing a magic constant---and checking whether the test
suite ``kills'' these slightly altered
versions of the original program by having at least one previously succeeding test fail. The ratio of
how many mutants the test suite kills onto how many we generate is
called the mutation score, ranging between 0 (no effective test) to 1
(the tests killed all mutants). Shy of merely telling whether a
statement has been visited or not, many researchers and practitioners
argue that the mutation score as a metric is superior to traditional
code coverage, as it truly exercises a program's behavior
\cite{gopinath2014code,inozemtseva2014coverage}.

However, this traditional approach to mutation testing is problematic in two aspects: First, with more than 100 known principled ways to mutate source
code, the space of mutated programs soon becomes prohibitively
large to compute~\cite{wong1995reducing}, let alone test, even
for very small programs. It is no surprise then that most studies on
mutation testing have been largely academic, a posteriori, come at
large computational costs, or been performed almost exclusively on
small projects~\cite{mutation_testing_survey,papadakis2019mutation,usaola2010mutation}. Second, it is unclear how a developer should best 
achieve an increase in mutation score---and whether such an increase has practical benefits, besides increasing coverage metrics~\cite{smith2009guiding}.

\begin{figure}[tb]
\centering
\includegraphics[width=\columnwidth]{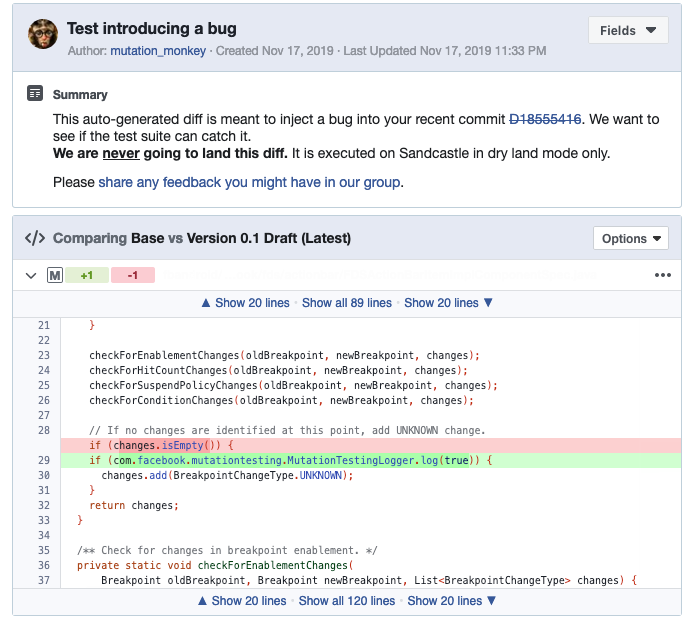}
\caption{Exemplary mutant created by Mutation Monkey.}
\label{fig:samplediff}
\vspace{-0.3cm}
\end{figure}

To address these problems, in this paper, we envision, implement, and evaluate an approach to make
mutation testing 1) feasible at scale in practice and 2) actionable to the individual developer under the name of ``Mutation Monkey.'' At Facebook's
scale, we have to be diligent with which mutants
we create, since we should not create mutants that would likely be
killed by the test suite (for the reason of computational expense of running
the test suite), or give no actionable signal to
developers (for the reason of conserving developer time). 
We thus empirically learn
error-inducing patterns from a corpus of common Java coding errors as
well as from changes that caused operational anomalies at Facebook
specifically.  To the best of our knowledge, this marks the first time that learned mutation operators have been used in industry. 

Building on Petrović's and Ivanković's 
work on surfacing an individual mutant~\cite{petrovic2018state}, we then
show developers not only which mutation slipped through the
test suite, as exemplified in \Cref{fig:samplediff} by making the {\tt if} condition always true; we also give them information on which tests visited, but failed to kill a mutation. Developers could use this as a suggestion for where and how to add a new 
or adapt an existing unit test to kill the mutation. 
We devised here a light-weight reverse test coverage logging
infrastructure for Java at Facebook that transparently wraps around the mutations to
provide real-time coverage information during test execution.
To
the best of our knowledge, this marks the first time these two normally opposed
concepts~\cite{just2014mutants,intro2019} have been
combined. 

In this paper, we perform two empirical case studies. In the first study, based on a two-month long execution of our mutation pipeline in practice, we assess

\textbf{RQ1:} How effective is the Mutation Monkey pipeline?

In the second study, we perform a randomized controlled trial with 26
software engineers at Facebook, in which we want to understand

\textbf{RQ2:} To which degree can location-based mutation testing
incentivize the creation of developer tests?

In summary, in a period of two months, Mutation Monkey created over 15,000 mutants. Our results show that all of the 16 learned mutation operators had a similar survival rate of around 60--70\%, which is significantly higher than the survival rate of generic mutation operators reported in industry (15\% at Google~\cite{petrovic2018state}) and in academia (often lower than 10\%~\cite{Ahmed2010,Cachia2013}). This indicates that \emph{using learned and arguably more realistic mutation operators makes mutation testing much more efficient and usable at scale}, since fewer compute resources are wasted for the generation and execution of easily killed mutants. Further, the user study indicates that \emph{almost all developers found the information of an identified test gap helpful} in principle. Only two saw no value in it, citing the infinite ways one could modify a piece of code. Moreover, where we could point to existing test cases covering a piece of code, this information was often instrumental for developers in deciding whether to add new tests, and helped them in finding the right place to do so. While developers would already act in almost half of the presented mutants by creating or adapting a test case, we find that the largest inhibitor to this is external knowledge about the code, e.g., that this piece of code would soon be deprecated. It is a challenge how to integrate such information to further increase actionability. Overall, though, the results demonstrate that Mutation Monkey can be successful in making mutation testing actionable by driving developers to create or adapt test cases.

This paper makes the following contributions:
\begin{itemize}
    \item Pioneering semi-automatic learning of mutation operators based on common Java bug patterns and Facebook-specific anomalies to make mutants more realistic.
    \item The first industrial implementation and deployment of this technique at Facebook under the name of \emph{Mutation Monkey} for its mobile code base (\Cref{sec:mutationmonkey}).
    \item An empirical study on the performance of Mutation Monkey, split into a quantitative study on the performance of mutation testing (\Cref{sec:eval}) and a user study part involving 26 developers at Facebook (\Cref{sec:user}).
\end{itemize}


\section{Background}
\subsection{Literature}
Mutation Testing has been widely studied since its inception in the 1970s~\cite{mutation_testing_survey,demillo1978hints,hamlet1977testing}. In this section, we describe the most relevant subset of the vast body of literature on mutation testing and how it relates to this article.

There are many works on reducing the costs of mutation testing~\cite{wong1995reducing,offutt2001mutation,just2017inferring,barbosa2001toward,frankl1997all}. 
For example, Just et al. investigated whether mutants can serve as a valid substitute for real faults in a large empirical study~\cite{just2014mutants}. They found that some traditional mutant operators were closer in resembling real faults than others. While their assessment at large was positive, they also found that traditional mutation operators do not expose certain types of real-world faults. In short, the above articles feature mainly academic case studies which aim to achieve a maximally high coverage while maintaining the smallest number of mutants or mutation operators. Our aim is different in that we want to present only the most relevant mutation to a developer to make it actionable to them. In contrast to the existing approaches that rely mostly on the more or less random generation of mutants, we do not have to filter a large amount of equivalent mutations due to our pattern-based approach---the patterns are largely non-equivalent by construction.

By contrast, the practical work by Petrović and Ivanković at Google is perhaps closest to ours~\cite{petrovic2018state}: They, too, integrated a scalable mutation analysis framework into the code review process. Similarly, we display the code changes Mutation Monkey generates via Facebook's code review tool, Phabricator~\cite{phabricator} (see \Cref{fig:samplediff}). Google reduces the computation amount by suppressing the generation of mutants for so-called arid nodes in the abstract syntax tree (AST, \cite{neamtiu2005understanding}), i.e., nodes and their transitive closure that are uninteresting to mutate, e.g., logging statements. Developers can further influence whether a given mutation might be uninteresting by giving feedback on it. While it is unclear which percentage of mutants lead to a code action on the developers' part (creating or modifying a test case), Petrović and Ivanković reported an average usefulness rate of 75\%. Instead of Google's manual process, we semi-automatically \emph{learn} only interesting mutation operators complete with context in which to apply them. Consequently, Mutation Monkey's mutant operators are more complex and numerous than the five that Petrović and Ivanković used in their study. They also reported having generated and tested 1.1 million mutants and surfaced 150,000 mutants to developers, an order of magnitude more than in this paper. However, it also shows that around 90\% of mutants at Google are potentially generated and tested wastefully.   

Building on a similar realization, Bingham Brown et al.~\cite{brown2017care} performed work in which they derived mutations from developer changes mined through projects' version histories. They show that so-called wild-caught mutants capture faults that traditional mutation operators are unable to and are at least as effective as traditional mutation operators. Like Mutation Monkey, Bingham Brown et al. reverse what they make out to be bug-fixing changes to obtain mutation operators. However, they controversially assume that any small change is a bug fix. We, on the other hand, mine our changes from a corpus of proven-to-be faulty changes: common fault patterns and bug fixes to changes that caused outages at Facebook. Mutation Monkey therefore has a realistic fault data set and, due to its learning capabilities, is not artificially constricted to very small changes. Changes can, in principle, be arbitrarily complex and long. In addition, by performing automated clustering and working on the abstract syntax trees (AST) of changes, as well as capturing the code context in which a bug was inserted, Mutation Monkey's pattern learning and application strategy is, we argue, more sophisticated.

In our work, the developer is the final arbiter of which mutant leads to the adaptation of an existing or the creation of a new test case. However, some developers expressed the wish to have this automated. In their visionary work, Fraser and Zeller used the information available from mutation testing to generate unit test cases automatically~\cite{fraser2011mutation}. However, in a real world system, the decision of whether, where, and how to adapt or create a test is complex and requires a holistic understanding of the test goal and strategy of the project~\cite{spillner2014software}. 

\subsection{Tools and Processes at Facebook}
\subsubsection{Getafix}
\label{sec:getafix}
Mutation Monkey relies on Getafix as its core component. Getafix~\cite{getafix} is a tool developed at and used by Facebook to automatically learn and apply code fixes to bugs uncovered by static analysis. When a developer at Facebook encounters certain static analyzer warnings during code review, e.g., for a null pointer, these warnings will come with an automated code change suggestion from Getafix to fix the flagged warning. 

Using Getafix starts with a training phase. In it, we show Getafix a set of file pairs, parsed to ASTs. Each such pair contains the AST of a piece of code before and after a bug fix. Getafix then uses a tree-differencing algorithm to map AST nodes between the before and after ASTs. For each pair of nodes mapped, if there is any difference between the subtrees rooted at those nodes, these subtrees are selected to comprise a concrete instance of an edit. The matched subtrees for which there is no difference are discarded, allowing Getafix to focus on the changed part of the code only.
Looking at the concrete edits of all set of file pairs, Getafix then uses a hierarchical clustering approach, combined with anti-unification, to find frequent tree structure patterns in all the before/after AST pairs it was given for training.  This process is adept at extracting bug fixing patterns at varying levels of granularity, which enables Getafix to learn fixes for a wide range of bugs even in the presence of additional, unrelated edits.  Tuning parameters allow the user to steer how close patterns have to be to each other for Getafix to cluster them together.

To generate a fix candidate with a before/after AST edit pattern, a buggy AST is scanned with the before AST from an edit pattern to find all matches, and each matched location is automatically transformed into the after AST. Once the fix is applied, we rank fixes by likelihood of being correct. For the top few fixes, we run the static analyzer that flagged a warning for the particular piece of code again to ensure that, after the application of the after AST edit pattern, the warning has in fact disappeared.

\subsubsection{Change-based testing}

\begin{figure}[tb]
\centering
\includegraphics[width=\columnwidth]{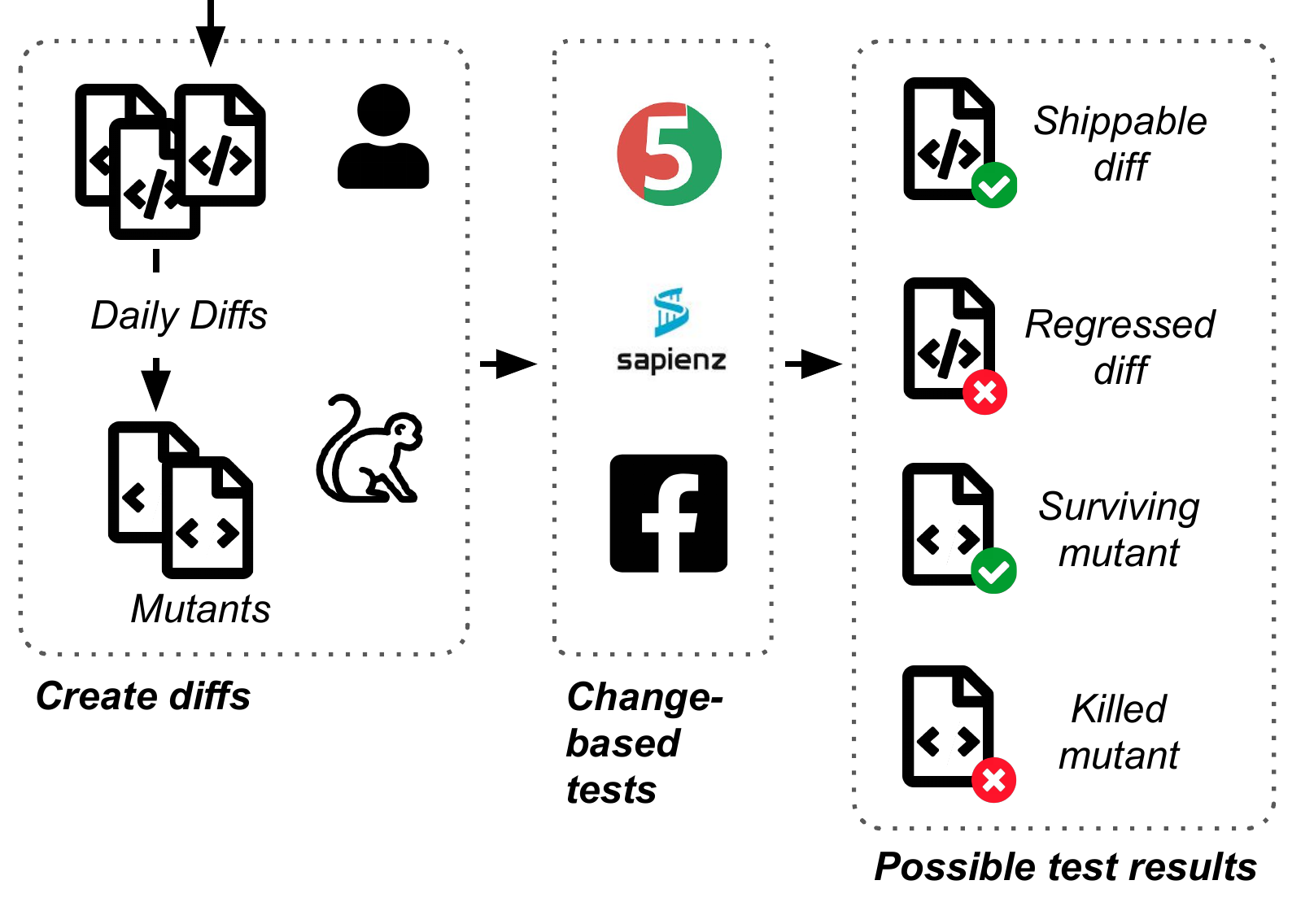}
\vspace{-0.4cm}
\caption{Both regular daily code changes by developers and mutants by Mutation Monkey are subject to the change-based testing process.}
\label{fig:change-based_testing}
\vspace{-0.5cm}
\end{figure}

Facebook maintains a comprehensive suite of unit, integration, and end-to-end tests for its mobile code base. Virtually all of these tests are eligible to run on changes submitted by developers at code review time, illustrated in~\Cref{fig:change-based_testing}. To reduce testing load, upon submission of a code change, a machine-learned, predictive test selection strategy chooses a subset of tests to exercise, maintaining at least 99\% chances of selecting a test that would have detected a regression~\cite{machalica2019predictive}. In addition, there is a cascade of further automated quality checks in place, ranging from automated static analysis tools~\cite{beller2016analyzing} to a search-based, automated testing approach called Sapienz~\cite{alshahwan2018deploying}. It is in this environment that we have to integrate \emph{Mutation Monkey.} Similar to how developers can only ship a diff if all relevant checks passed, as \Cref{fig:change-based_testing} shows, we only surface surviving mutants to developers.

To scale mutation testing at Facebook, we time submission of mutants such that they are tested outside of peak hours. Systems supporting testing changes submitted by developers typically see lower use between 7 p.m. and 7 a.m. Pacific time, which lets us evaluate hundreds to thousands of mutants per day without incurring noticeable infrastructure cost.

\section{Mutation Monkey}
\label{sec:mutationmonkey}

\begin{figure*}
    \centering
    \subfigure[Mutation operator learning pipeline.]
    {
        \includegraphics[width=\textwidth]{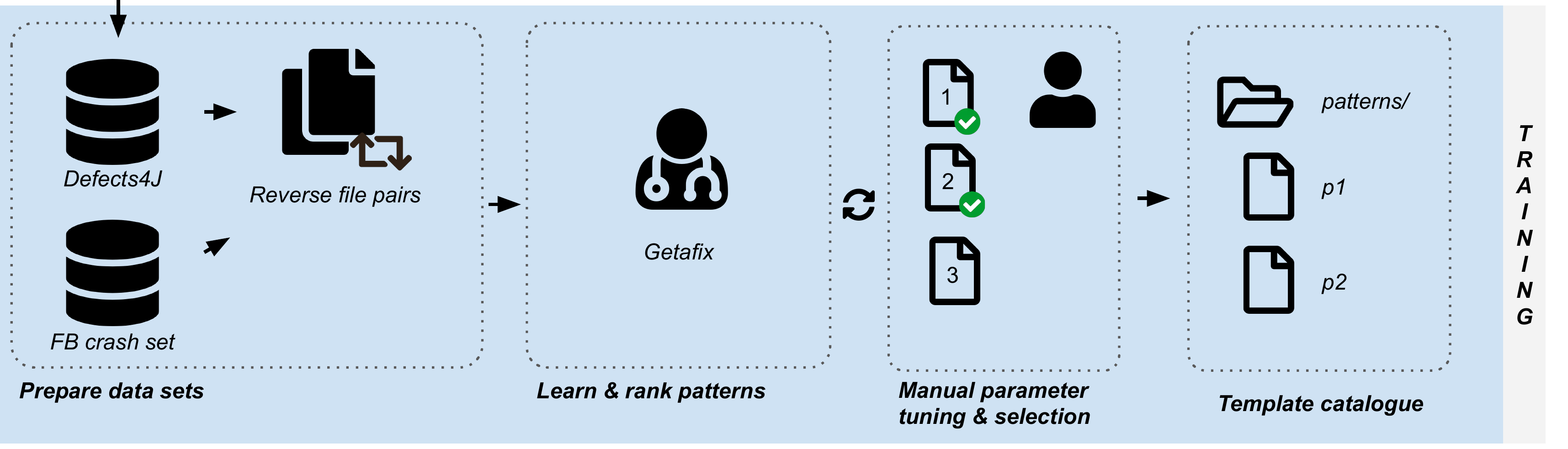}
        \label{fig:pipline-learning}
    }
    \\
    \subfigure[Mutant creation/template application pipeline.]
    {
        \includegraphics[width=\textwidth]{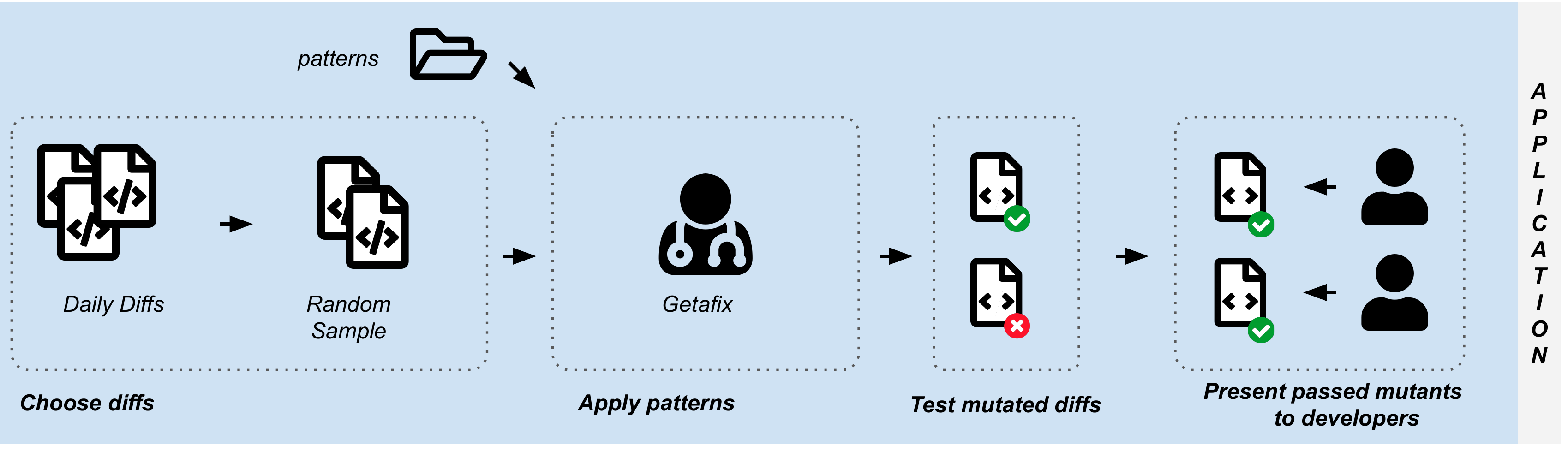}
        \label{fig:pipeline-application}
    }
    \caption{Overview of the implementation of mutation testing at Facebook.}
    \label{fig:pipeline}
    \vspace{-0.2cm}
\end{figure*}

In this section, we describe how Mutation Monkey works and is implemented at Facebook, outlined in \Cref{fig:pipeline}. There are two separate processes, the training (top) and application (bottom) pipelines of Mutation Monkey.

\subsection{Training: Semi-automatically learning new mutation operators from past changes}
\label{sec:learning-mutation-operators}
\begin{figure*}[tb]
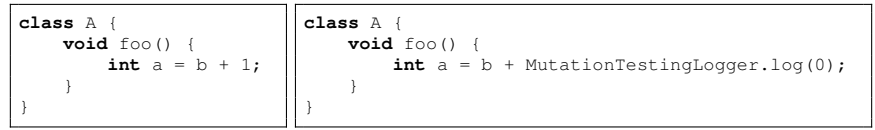

  \centering
  \scriptsize
  \begin{minipage}[t]{0.19\linewidth}
\begin{lstlisting}[language=Java,frame=single]
class A {
    void foo() {
        int a = b + 1;
    }
}
\end{lstlisting}
\end{minipage}
~~
\begin{minipage}[t]{0.41\linewidth}
\begin{lstlisting}[language=Java,frame=single]
class A {
    void foo() {
        int a = b + MutationTestingLogger.log(0);
    }
}
\end{lstlisting}
\end{minipage}

\vspace{-0.1cm}
\caption{Exemplary learning pair for Getafix and pattern {\tt LITERAL\_TO\_ZERO}.}
\label{lst:learning_pair}
\vspace{-0.2cm}
\end{figure*}

\Cref{fig:pipline-learning} gives a graphical overview of how we learn mutants from faulty changes.

In the past, construction of mutation operators was mostly arbitrary: Researchers and developers envisioned small changes to source code that were intended to somehow change its behavior and then implemented them as templates. A typical example is changing a ``+'' to a ``-'' operator. This turned out to produce many mutants that either did not produce syntactically correct programs in the first place or that were easily killed by the most simple of tests. 

Instead, we argue that by learning mutants from a corpus of previous real faults, we can produce more realistic---and thus, more actionable and harder to kill---mutants. Instead of fixing bugs, we use Getafix to place defects into the code. As the patterns Getafix learns consist merely of a before and an after AST, Getafix itself has no notion of what a bug-fixing and what a bug-introducing pattern is; going from a bug-fixing pattern to a bug-introducing pattern is simply done by \emph{changing the before and after pairs in its training phase} or reversing the direction of its pattern application (both are equivalent to each other). For Mutation Monkey, we use Getafix to learn bug-inducing patterns from several different data sources:

\subsubsection{Defects4J}

Defects4J is a collection of reproducible bugs extracted from highly used Open Source Software projects written in Java~\cite{just2014defects4j}. At the time of our investigation, we used the latest available version 1.4.0 of the data set. From Defects4J, we obtained 516 file pairs of 438 random and real bug fixes from 6 open-source Java projects. In total, we learned 11 bug-inducing patterns from Defects4J.

\subsubsection{Production crashes}
Facebook tracks and maintains a database of crashes that happen in production. If there was a code patch associated with fixing such a crash, it is linked in the according task. In contrast to \emph{Defects4J}, we do not directly obtain bug-inducing code changes this way. However, conceivably, by reversing a fix to a crash, we can (re-)introduce a crash, or at least an important part of a situation that lead to the crash. We surmise that armed with such evidence, instead of a fuzzy increase in mutation score, developers would be more willing to implement a test case that prevents such a scenario. Consequently, we mine an archive of crash fixes, reverse their direction, and let \emph{Getafix} learn their patterns. \emph{Getafix} then produces a list of patterns ranked by frequency of occurrence. In line with previous research~\cite{osman2014mining}, the most highly ranked such pattern was the removal of a {\tt null} check. We compiled a data set of more than 18,000 code fixes to crashes in Facebook's mobile code base, and extracted 4 patterns from them that were not already contained in the Defects4J data set. 

\subsubsection{Test failures}
In addition to past production crashes, we compiled a second data set at Facebook comprising of small modifications that made an originally failing test pass, and vice versa. We mined this data set by looking at all commits (henceforth referred to as ``diffs'') in the mobile code base that operated on Java code and that originally had at least one test fail on them. Then, the author made some a modification that lead to the same test passing (or vice versa). We mined 179 file pairs that flip test results from failed to passed (i.e., bug fixing changes) and 175 pairs that flip test results from passed to failed (i.e., code breaking changes). We synthesized 5 patterns from them.

Combining both Facebook-specific data sets, we extracted 7 unique bug-introducing patterns.
In addition, we added a pattern introducing a null dereference manually, since we suspected it to be frequently applicable in practice, given prior domain knowledge. We ended up with 19 patterns in total.

While learning, ranking, and generalizing of patterns from these data sets is automatic, it does involve hyper-parameter tuning of Getafix's learning engine. It also requires expert assessment to choose which and how many patterns to implement. Finally, there is  manual work involved in converting the patterns to mutation templates, e.g., to insert a wrapping call to the test logging infrastructure, as \Cref{lst:learning_pair} shows. This is why we call the learning semi-automatic.


\subsection{Application: Automatically applying the learned mutation operators to generate mutants}
\label{sec:apply}

\Cref{fig:pipeline-application} gives a graphical overview of how we apply the learned mutant templates to production code.

From an operational standpoint, the only difference in Mutation Monkey's to Getafix's  application phase is that Getafix normally applies its bug-fixing patterns based on the signals of a static analysis warning. However, it has no such warning information to seed its mutations into for mutation testing because we place them into fully working code. Therefore, there are potentially many more places to seed mutations into than where bug-fixing patterns would apply. Similarly to Google, we avoid applying mutants in certain unprofitable spots, such as in direct logging calls~\cite{petrovic2018state}. Finally, the application of a mutation does not necessarily guarantee the creation of syntactically valid mutant program. To catch such faulty programs early, we run a light-weight syntax checker immediately after the generation of a mutant and before its submission as a diff.

To determine which, if any, of the automated tests kill a particular mutant by \emph{Mutation Monkey}, we submit the mutant as a code change to the code review system. \Cref{fig:change-based_testing} illustrates how both mutants and code changes made by developers are subject to the same change-based testing process at Facebook and subject to all validation methods available at our disposal. We await completion of the validation process and use information typically shown to the author of the change to determine whether the mutant has been detected and by which test. In the end, we only present unkilled mutants.

\subsection{Test Location Logging}
In addition to making the mutation itself, we also insert a logging statement around the mutation. This is so that we can get information about which tests visited the mutated statement or block how many times. Our intuition was that armed with such knowledge, developers could more easily (1) make the right decision of whether to test a given piece of code and if so, (2) identify where to adapt or place such a new test.

\section{Quantitative Evaluation}
\label{sec:eval}
In our first empirical case study, we assess:

\textbf{RQ1:} How effective is the Mutation Monkey pipeline?

\subsection{Study Design \& Methods}
\label{sec:pipstudydesign}
To answer this question, we first derived a set of learned mutation operators, then integrated their application and evaluation in the daily testing and validation process at Facebook under the name ``Mutation Monkey'' (see \Cref{sec:mutationmonkey}). 

Among all diffs developers created in Facebook's mobile code base in the past day, Mutation Monkey selects 100 diffs touching its {\tt fbandroid} platform at random. It filters out any changes that do not occur in Java classes. Then, it sequentially tries to apply mutation operators on the changed Java lines in the given diff, until the first template applies successfully. Since, e.g., not all diffs touch an {\tt if} statement, it is easy to see why not all patterns apply everywhere. This study design makes it harder to cross-compare different operators, but it allows us to draw conclusions on how many times a pattern can occur in practice, which is an important signal for the creation and maintenance of mutation operators.

\subsection{Results}

\begin{figure*}[tb]
\centering
\includegraphics[width=2\columnwidth]{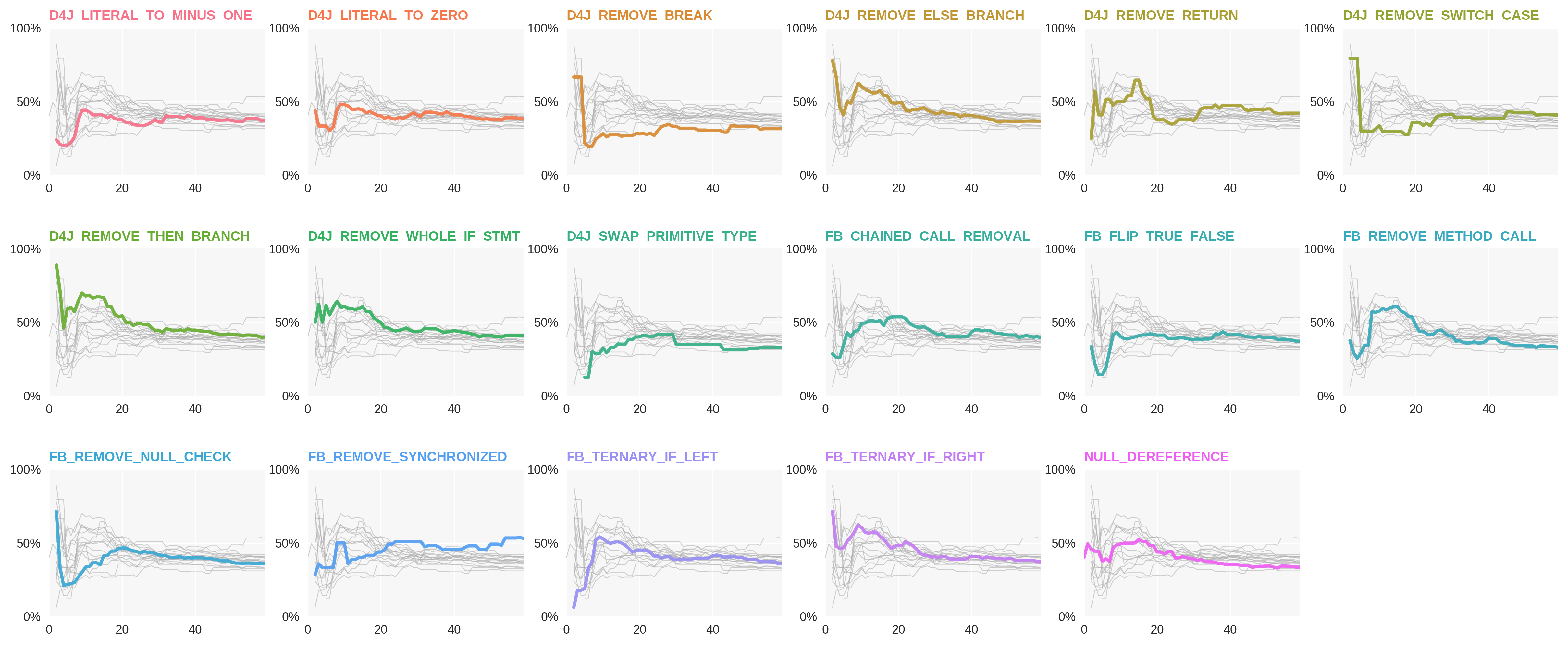}
\caption{Expanding average mutation kill rate over time for mutation operators with at least 100 mutants.}
\label{fig:killrate_over_time}
\end{figure*}

\begin{table*}[tbh!]
\caption{Mutation operator statistics for operators with at least 100 mutants.}

\resizebox{\linewidth}{!}{

\begin{tabular}{llrrl}
\toprule
\textbf{Mutation template name} & \textbf{Source} & \multicolumn{1}{l}{\textbf{Occurrence}} & \multicolumn{1}{l}{\textbf{Kill rate}} & \textbf{Template (simplified)} \\
\midrule
{\tt LITERAL\_TO\_MINUS\_ONE}    & D4J             & 1,161                                    & 37.0\%             & {\tt 1} $\xrightarrow{}$ {\tt -1}                               \\
{\tt LITERAL\_TO\_ZERO}          & D4J             & 1,154                                    & 38.1\%             &  {\tt x} $\xrightarrow{}$ {\tt 0}                              \\
{\tt REMOVE\_BREAK}              & D4J             & 1,054                                     & 30.6\%             &   {\tt case 1: A(); break; case 2: B();} $\xrightarrow{}$ {\tt case 1: A(); case 2: B();}                             \\
{\tt REMOVE\_ELSE\_BRANCH}       & D4J             & 1,054                                     & 35.0\%             &   {\tt else \{ ... \}} $\xrightarrow{}$ {\tt else \{;\}}                             \\
{\tt REMOVE\_RETURN}             & D4J             & 493                                     & 40.3\%             &   {\tt return;} $\xrightarrow{}$ {\tt ;}                             \\
{\tt REMOVE\_SWITCH\_CASE}       & D4J             & 754                                     & 40.8\%             &   {\tt case 1: A(); break; case 2: B();} $\xrightarrow{}$ {\tt case 1: A(); B();}                             \\
{\tt REMOVE\_THEN\_BRANCH}       & D4J             & 1,030                                     & 39.9\%             &   {\tt if(A) \{ ... \}} $\xrightarrow{}$ {\tt if(A) \{;\}}                             \\
{\tt REMOVE\_WHOLE\_IF\_STMT}    & D4J             & 1,143                                    & 40.0\%             &  {\tt if(A) \{ ... \} else if(B) \{ ... \} else \{ C(); \} } $\xrightarrow{}$ {\tt ;}                              \\
{\tt SWAP\_PRIMITIVE\_TYPE}      & D4J             & 167                                     & 32.3\%             &   {\tt double a; $\xrightarrow{}$ {\tt int a; }}                             \\
{\tt CHAINED\_CALL\_REMOVAL }      & FB              & 1,042                                    & 39.4\%             &  {\tt a.b(1).b(2).c(); } $\xrightarrow{}$ {\tt a.b(1).c(); }                              \\
{\tt FLIP\_TRUE\_FALSE}           & FB              & 967                                     & 36.9\%             &   {\tt if(a == true)} $\xrightarrow{}$ {\tt if(a == false)}                             \\
{\tt REMOVE\_METHOD\_CALL}        & FB              & 1,279                                    & 32.8\%             &  {\tt a();} $\xrightarrow{}$ {\tt ;}                              \\
{\tt REMOVE\_NULL\_CHECK}         & FB              & 808                                     & 34.9\%             &  {\tt if(variable == null) \{ ... \}} $\xrightarrow{}$ {\tt ;}                           \\
{\tt REMOVE\_SYNCHRONIZED}        & FB              & 143                                     & 53.1\%             &  {\tt synchronized Object foo() { ... }} $\xrightarrow{}$ {\tt Object foo() { ... }}                              \\
{\tt TERNARY\_IF\_LEFT}           & FB              & 677                                     & 36.2\%             &  {\tt a ? b : c} $\xrightarrow{}$ {\tt b}                            \\
{\tt TERNARY\_IF\_RIGHT}          & FB              & 659                                     & 33.5\%             &  {\tt a ? b : c} $\xrightarrow{}$ {\tt c}                              \\
{\tt NULL\_DEREFERENCE}               & -               & 1,908                                    & 38.6\%             &  {\tt String s;} $\xrightarrow{}$ {\tt @javax.annotation.Nullable String s; s.toString();} \\
\midrule 
$\Sigma$ & 2 & 15,493 & - \\
\bottomrule 
\end{tabular}
}
\label{tab:killrate}
\end{table*}

\begin{table*}[tb]
\caption{Descriptive results of the mutation coverage logging.}

\resizebox{\linewidth}{!}{

\begin{tabular}{lllll}
\toprule
\# Unique test cases & \# Visits of mutation in test cases & \# Test case visits & \# Tests case visits for surviving mutants & \# Test case visits for killed mutants \\
\midrule   
2,060 & 3,586,793 & 15,522 & 3,612 & 11,910 \\
\bottomrule 
\end{tabular}
}
\label{tab:coverage}
\end{table*}

\Cref{tab:killrate} shows the statistics of various mutation templates. While kill rates are quite similar across mutation operators, there are large discrepancies in how many times a mutation template applied successfully. There were two operators with fewer than 100 applications ({\tt REMOVE\_EXPLICIT\_CAST} and {\tt FOR\_OFF\_BY\_ONE}), which we do not report on here for statistical reasons. For example, the pattern {\tt NULL\_DEREFERENCE} was applicable 13 times more than the least applicable operator, {\tt REMOVE\_SYNCHRONIZED.}   
Interestingly, the pattern {\tt REMOVE\_SYNCHRONIZED} at 53\% was also the most frequently discovered by tests.

\Cref{fig:killrate_over_time} shows the expanding average mutation kill rate over time for each of the 17 mutation operator with at least 100 applied mutants. This means that each sub-plot depicts the kill rate in \% (vertical axis) over the study time expressed in days (horizontal axis) for the specific mutation operator annotated above it. For example, for day 3 on the horizontal axis, the first sub-plot shows the average kill rate of all mutants generated based on the {\tt D4J\_LITERAL\_TO\_MINUS\_ONE} template up to day 3, at around 20\%. The inherent jitter in the sub-plots stems from the fact that the Mutation Monkey pipeline creates a variable number of mutations based on the diffs of that day (see \Cref{sec:pipstudydesign}). Similarly, the well-testedness of the diffs in which a mutation happens to apply can bias a mutation operator's kill rate when looking at a short time sample. To cross-compare mutation operator performance, it is therefore important to consider a sufficiently large sample. \Cref{fig:killrate_over_time} shows that (1) individual day influences peter out around the 30 day mark and that (2) most mutation operators' kill rate stabilizes between 30\% to 40\%, indicating that there seem to be no large differences in the kill rates of mutation operators.

\Cref{tab:coverage} presents descriptive statistics on test coverage. Overall, all mutations were visited more than 3.5 million times (this includes cases where the mutation is recursively called or inside a loop).  A set of 2,060 unique tests triggered these visits. The table also shows the diverging number of visits for surviving and killed mutants.

\section{User Study}
\label{sec:user}

In our second study, we perform a randomized controlled trial with 26
software engineers at Facebook, in which we want to understand

\textbf{RQ2:} To which degree can location-based mutation testing
incentivize the creation of developer tests?

\subsection{Study Design \& Methods}
\label{sec:interviedesign}

\begin{figure*}[tb]
{\footnotesize
\begin{framed}
Line 1: Hey, Alisha!\\	
Line 2: I saw you recently committed D18848361.	I did a tiny, buggy modification to your original diff in D18873388 (focus on the small actual code change, ignore the logging around it and the BUCK file). I wanted to see if the tests in {\tt fbandroid} catch this slight change, that is, at least one test should fail on this buggy version. My 'faulty change' was based off of a set of common general and {\tt fbandroid}-specific Java defect patterns.\\
Line 3: The thing is: none of the existing tests (e2e, integration, unit) was able to catch this faulty change. In your opinion, does D18873388 possibly expose a lacking testing? :) \\
Line X: Were you aware that there are Y different tests in {\tt fbandroid/} that covered this particular mutation, but failed to kill it?
\end{framed}
}
\vspace{-0.5cm}
\caption{Interview protocol with pre-defined text snippets.}
\label{fig:interview}
\end{figure*}

From November 23rd until December 20th 2019, we conducted a user study in which we approached 29 developers and showed them an unkilled mutant. 

The interviews were semi-structured, remote (via Facebook's internal messaging system), and interactive, although the initiation could start asynchronously. We did this in an iterative way to refine the study protocol after the first three interviews, which we do not report on. \Cref{fig:interview} shows the final version of the initiating messages we approached 29 developer with. Our goal was to assess whether 1) our mutants indeed revealed a possibly missing test, 2) developers would act on them (and even more importantly, if not, why not,) and 3) whether surfacing reverse test coverage for the precise mutation would be useful.

We prepared our study sample so it comprises 50\% mutants with and 50\% mutants without coverage information. We then contacted the author of each of the 29 diffs. If there was a diff that for some reason was unsuited for mutation, we picked another already generated diff for the same author. From the 29 diffs, we received suitable 26 answers (90\% response rate). All but one respondent were software engineers in different teams (the one other respondent was a Manager). Their average tenure at Facebook was 2 years and 10 months, with a maximum of 7.6 years, and a minimum of 7 months. 

After leading the interviews, we categorized them. Specifically, we set out to answer the five questions in~\Cref{tab:interviews}.  Each interview should yield a ternary response to each of the questions, unless it was aborted early (e.g., in case the interviewee no longer responded). To categorize the interviews as unbiased as possible, the lead author of the study interpreted them with a time gap of half a year after doing them. In addition, a second author independently went through the interviews to rate them along each of the axes without further instructions. Then, both authors discussed their findings. While initial agreement was quite high, we noticed that the second author had a much higher use of the ``unclear'' category. In a grounded theory-inspired process, both authors went through their differences together. This changed some obvious miss-categorization on both authors' assessments. In addition, some of the ``unclear'' ratings could be cleared up. In total, the first author changed 23 ratings (17.7\% of all ratings) and the second 19 ratings (14.6\%). This yielded an interrater reliability agreement of $\kappa=1-\frac{1-p_0}{1-p_e} =1-\frac{1-\frac{115}{115+15}}{1-\frac{1}{130^2}\cdot(62\cdot52+13\cdot12+55\cdot66)}=1-\frac{1-0.884}{1-0.414}=0.802.$ According to the interpretation guidelines by Fleiss et al.~\cite{fleiss1981measurement}, this is considered ``excellent.''

\subsection{Results}
\begin{figure*}[tb]
{\footnotesize
\begin{framed}
P2: ``But there is drawback here, since you actually can see, or find out the missing unit test. Instead of providing those diff to point out the potential unit test, why not just add those unit test? Unless you are able to generate those checking diffs automatically.'' \\
P4: ``but still that will just change the number that being sent to backend logging.''\\
P5: ``Oh really? I didn’t know that lol.''\\
P9: ``So if I added screenshot test at the very beginning, I will get a lot of test failure tests.''\\
P19: ``Are you building some cool bot that will add test automatically?'' \\
P20: ``In this [...] area of code, we've gone to great lengths to get people to write more unit tests. But, there are still [...] gaps, and I think that's okay.'' \\
P25: ``Yes. And I'm a bit disappointed that our e2e tests didn't catch that.'' \\
P25: ``If this weren't deprecated I'd create a task to fix it tomorrow. But it's deprecated, so for now I won't worry about it.'' \\
P26: ``[Mutation Monkey] seems like a useful way to prevent those overlookable bugs that really easily get commited.''
\end{framed}
}
\vspace{-0.5cm}
\caption{Quotes by participants.}
\vspace{-0.3cm}
\label{fig:quotes}
\end{figure*}

\begin{table*}[tb]
\caption{Descriptive results of the user study with 26 developers.}

\centering
\resizebox{0.65\linewidth}{!}{

\begin{tabular}{lrrr}
\toprule
Question & Agreed & Disagreed & Unclear/NA \\
\midrule
Does the diff expose lack of testing? & 84.6\% (22/26) &  7.7\% (2/26)  & 7.7\% (2/26) \\
Are such diffs helpful? & 61.5\% (16/26)   & 0\% (0/26) & 38.4\% (10/26)   \\
Are you going to add a test? &  46.2\% (12/26) & 23.1\% (6/26)  & 30.8\% (8/26) \\
Was coverage information new? &  26.9\% (7/26) & 15.3\% (4/26)  & 57.7\% (15/26) \\
Was coverage information helpful? & 19.2\% (5/26) & 6.3\% (1/26) & 76.9\% (20/26)\\
\bottomrule 
\end{tabular}
}
\label{tab:interviews}
\vspace{-0.1cm}
\end{table*}

\Cref{tab:interviews} presents the assessment results of the interviews from the first author in an aggregated format. Not applicable summarizes all categories where we either did not ask this question, interviewees did not respond to it, or the flow of the conversation made asking the question redundant. In addition, we extracted one or two insightful quotes from participants and present them verbatim in~\Cref{fig:quotes}. In the following, we reference the quotes in the figure via the participant code PX.

The general impression from the interviews was that most---if not all---developers had not heard of the concept of mutation testing before. After our explanations, close to 85\% of interviewees found Mutation Monkey to be a generally valuable tool that could support them in their testing efforts (P26). 

When asked whether they would write a test for the specific gap we found, almost half (11) confirmed to do so (42.3\%). This included three cases in which developers submitted a diff with tests independently from us after their original diff, but before we interviewed them.

Developers gave a plethora of reasons for not writing a test based on the mutants: Some wanted us to come up directly with the test instead of just the test hole (P2, P19), similar to Fraser and Zeller~\cite{fraser2011mutation}. Some indicated that we modified code that is of minor importance, e.g., modifying the value of a variable that is only used for logging (P4). The simple approach to filtering out logging lines (see \Cref{sec:apply}) was not enough in this case. In the future, we could compute the data flow or AST to check for such cases. One developer wrote that they would add a test immediately if the code was not about to be deprecated, a fact not visible from the code (P25). Some expressed that this was new code likely to undergo iteration before stabilizing, others that they do not have time to write a test for it, and others still that the state of testing for this part of the code base was simply lacking.

\section{Discussion}

\subsection{Interview Insights}
From the interviews, it became quickly clear that most developers had not known mutation testing beforehand. Therefore, the way we surface a mutant to them is important. All developers eventually understood the concept behind it, but in most cases, the simplified explanation in our mutant diff did not suffice. In addition, some developers were confused by the logging statement that we included in the mutation of the code to obtain coverage information; in such cases, we explicitly pointed out the actual mutation. Here, a better UI would allow us to hide the insertion of the logging information from developers ({\tt MutationTestingLogger.log} in \Cref{fig:samplediff}) and instead let them focus on the actual code mutation at hand ({\tt changes.isEmpty() $\xrightarrow{}$ true}).

We observed that often, there was no relationship between the introduction of a small mutation and its explicit need to be tested. As P20 indicates, there could be arbitrarily many ways to mutate a program in ways that do not warrant tests. We tried to come around this problem by basing the mutation operators on past crashes at Facebook and other common error-inducing patterns, but this did not convince all developers.

In serveral cases, developers pointed to tests they had written which would actually catch the mutant we presented to them. At the time when we selected the diff to mutate, these tests were not available yet. In a sense, this is a very strong validation of the value of the learned mutation operators. At the same time, it raises the question how we would find such cases and not give an unnecessary signal. As we learned from P9, adding tests early might overload developers with noise from failing tests, particularly in rapidly changing code.

\subsection{Coverage Information}
Coverage of mutations reflects the state of testing at Facebook. We can attribute the fact that only a relatively small set of 2,060 unique tests is responsible for all mutation visits to the prevalence of screenshot and UI testing in the mobile code base. Moreover, as expected, we see that there is a strong correlation between the number of test cases covering a mutation and its kill state: there are almost four times as many test case loggings for such mutants, even though they are almost as frequent as unkilled mutants (see \Cref{tab:killrate}).

While the results on the helpfulness on mutation coverage in \Cref{tab:interviews} may initially seem low, we only had coverage information available for half of the diffs we assessed in the user study to begin with (see \Cref{sec:interviedesign}). Moreover, since it was the last question, there could be many reasons why the interview ended prematurely. For the developers to which we could present this information, the majority appreciated it. Beyond what the numbers can tell, revealing this information seemed to make a large difference in the conversation flow (see, e.g., P5). Only one developer found the information not helpful because they were very familiar with the code and its associated tests.

\subsection{Putting Mutation Kill Rates in Perspective}
Comparing the results of mutation operators on different code bases is difficult, since they are inevitably tied to the thoroughness of the underlying test suite: A strong test suite might kill even the best of mutants, while a lax test suite might make one overrate the performance of the generated mutants. Hence, any comparison between mutant operators across programs must be consumed cautiously, even under the assumption that testing is at a similar maturity. In our case, we mutate production program code from Facebook's mobile code base, which is under extensive unit, integration, and system testing. As such, we assume that testing is as thorough as is feasible in practice. 
The mutant survival rates we reported in this paper are far higher than those typically reported for traditional mutation operators~\cite{Ahmed2010,Cachia2013}, in industry~\cite{petrovic2018state}, or even of wild-caught mutants~\cite{brown2017care} that somewhat resemble Mutation Monkey's. This is an indication that the mutants of Mutation Monkey are more adapted to its destination code base, have a higher specificity of finding real test holes and bugs than traditional mutation operators, and are far more complex than the one or two character in-place edits of traditional mutation operators. 

However, our conclusion is only evidence, not proof per se. We need more studies on other systems comparing a learned approach to previous, more traditional approaches of deriving mutation operators. For such a comparison, we would need an open benchmark stemming from a open-source system that not only features an industry-grade test suite (similar to Space~\cite{vokolos1998empirical,brown2017care}, frequently used in the evaluation of mutation operators because of its rigid test suite), but also a rich version control history along with a database of bug-inducing commits. 

\subsection{On The Value of Mutation Testing}

The number of occurrences of a mutation operator does not yet attribute a value to the potential bugs found by the operator; but it indicates that some mutation operators might be more profitable in their bug finding capabilities than others when applied to real-world changes. For mutation operators for which we found a difference in the kill rate, this could indicate that developers are aware of intrinsically hard-to debug bugs, for example when they add ``synchronized'' to a method, and therefore tend to test more for it. The reverse could be true for inconspicuously looking code fragments like the ternary operator: As P26 put it, ``those overlookable bugs that really easily get commited.'' Fetching these can be a selling point for mutation testing.

A bottom-line question on the value of mutation testing is: Is adding tests to kill surviving mutants a superior way to prevent future bugs to other criteria to add tests? This question is important because developers are generally already pressed for time, and writing additional tests has a cost.
Unfortunately, there is no easy way to answer it.

First, it is hard to estimate the likelihood that the mutants, if released, would actually cause observable defects. We perhaps could
do this estimation by releasing the unkilled mutants in the field and catch operational abnormalities, but this carries the risk of causing increased user dissatisfaction.  For this reason, we strive to establish only that mutation testing is an effective way to encourage writing tests developers consider useful, as opposed to a stronger result that mutation testing is an effective way to encourage writing tests \textit{that prevent real-world, costly regressions in production.}

Second, it is also hard to compare against (hypothetical) alternative strategies to spend testing budget
that are potentially more productive in preventing future bugs.  After all, the application has been in production
for a while, is reasonably stable, and the test suite has some level of statement coverage already.  We cannot
point easily to obvious holes in testing that would be the first thing to plug.

In the end, we rely on a ``social'' proof.  If the presentation of unkilled mutants---coupled with information that the mutants are representative of past bugs---motivates developers enough to write tests, this is a good outcome; even if it is not known to be superior to other hypothetical strategies to add tests.

\subsection{Threats to Validity}
In this section, we describe the three main threats we identified to our study and how we mitigated them.

\subsubsection{Data quantity} In this study, we only generated 15,493 mutants, compared to 322,972 by Brown et al.~\cite{brown2017care} and 1.1 million at Google~\cite{petrovic2018state}. The picture changes, though, when assessing (a) how many diffs we operate on (15,493 versus 77,000 at Google~\cite{petrovic2018state}) and (b) how many mutants could successfully be compiled (100\% versus only 12-15\%~\cite{brown2017care}), as all our mutants compile \emph{by construction} (see \Cref{sec:getafix}). Moreover, it was the explicit goal of this paper to reduce the computational effort required to do mutation testing by lowering the number of mutants to generate and evaluate. Lastly, \Cref{fig:killrate_over_time} shows that, after some initial up and downs caused by daily fluctuations, the performance of mutation kill operators stabilized toward their long-term kill rate, indicating that we have obtained enough data to make accurate statements. 

\subsubsection{Interviewee experience} Most interviewed engineers did not know mutation testing before the study, which could lead to misunderstanding and bias the results. Due to the interactive nature of interviews, we responded and explained the concept at the individually needed pace. We are confident that every interviewees developed a clear understanding of Mutation Monkey, making a control group of developers already familiar with mutation testing unnecessary.

\subsubsection{Flaky tests} Non-deterministic tests represent a concern in large software systems~\cite{bell2018deflaker,micco2016flaky}. Mutation testing is based on the assumption that if a test suite passed on the original program but fails on the mutated version, then this is due to the mutation. However, in the presence of flaky tests, this assumption no longer holds. By inserting a logging statement into the mutation, we can ensure that a test actually executed the mutated piece of code. We thus disregard failing test results from other parts of the system. We employed this---to the best of our knowledge---novel strategy, in addition to Facebook's general strategy of disabling persistently flaky tests, to mitigate the issue of flaky tests outside a mutation. 

\section{Future Work \& Conclusion}
In this paper, we pioneered the idea of \emph{learning} mutation operators. Naturally, there is room for future extensions:

\begin{itemize}
\item Not all developers found the argument that the mutations are mined from past crashes and common bug patterns convincing enough to act. They argued that there is often a near-infinite number of ways to violate the contract of a program, and it is not a good idea to write tests to exclude all of those.  An idea to directly measure the impact of a mutant, and better justify why it needs a test, is to canary it in practice, i.e., ship the mutants to a small percentage of users. If production metrics deteriorate, we might convince developers more easily of the need to test.

\item To provide developers with an idea of tests that already exist, we measured exactly which tests covered a mutation. In the future, it would be interesting to ``upsert'' the mutation logging statement into the surrounding body, method, and class. This way, we could obtain tests that did not exactly visit the mutation at hand, but came close. We surmise that such information (1) might be available for more mutants and (2) might increase the helpfulness rating of the coverage information.

\item Concrete actionability on mutants is the end metric that matters. The factors that lead to no action on the developers' side often had nothing to do with the mutation per se. Instead, they regarded mostly contextual information about a spot of code, e.g., that this is an area that does not need testing or that would be deprecated soon, and hence any testing effort would be wasted. If we could include such external information into Mutation Monkey, it would likely help increase actionability.
\end{itemize}

In summary, this paper marks the first time complex mutation operators have been learned from past bug-inducing changes and seeded in a large industrial system, demonstrating the scalibility of the approach. The high survival rates of the learned mutation operators indicates that they are better apt at finding test holes in the target system than most traditional operators. When faced with a mutant, the majority of developers found such information helpful in principle and almost half of developers were planning to take or had already taken action based on the identified test gap. Factors for not acting often lay outside the scope of Mutation Monkey. Finally, coverage information, where available, played an important role in helping developers understand a mutation.

\bibliographystyle{IEEEtran}
\tiny
\bibliography{IEEEabrv,paper}

\end{document}